\documentclass[journal]{IEEEtran}
\usepackage{cite}

\ifCLASSINFOpdf
   \usepackage[pdftex]{graphicx}
   \graphicspath{{../pdf/}{../jpeg/}}
   \DeclareGraphicsExtensions{.pdf,.jpeg,.png}
\else
   \usepackage[dvips]{graphicx}
   \graphicspath{{../eps/}}
   \DeclareGraphicsExtensions{.eps}
\fi
\usepackage{enumerate}
\usepackage{amsfonts}
\usepackage{booktabs}
\usepackage[cmex10]{amsmath}

\usepackage{mathabx}
\begin{document}
%
\title{Joint User Association and Interference Mitigation for D2D-Enabled Heterogeneous Cellular Networks}
%
%
%

\author{Tianqing~Zhou,~
        Yongming~Huang,~\IEEEmembership{Member,~IEEE,}
        and~Luxi~Yang,~\IEEEmembership{Member,~IEEE,}
\thanks{T. Zhou is with the School of Information Science and Engineering, Southeast University, Nanjing 210096, China, and also with the State Key Laboratory of Millimeter Waves, Department of Radio Engineering, Southeast University (e-mail:zhoutian930@163.com).}
\thanks{Y. Huang is with the School of Information Science and Engineering, Southeast University, Nanjing 210096, China, and also with the Key Laboratory of BroadbandWireless Communication and Sensor Network Technology (Nanjing University of Posts and Telecommunications), Ministry of Education (e-mail:huangym@seu.edu.cn)}
\thanks{L.Yang is with the School of Information Science and Engineering, Southeast University, Nanjing 210096, China (e-mail:lxyang@seu.edu.cn).}}

\maketitle

\begin{abstract}
The heterogeneous cellular networks (HCNs) with device-to-device (D2D) communications have been a promising solution to cost-efficient delivery of high data rates. A key challenge in such D2D-enabled HCNs is how to design an effective association scheme with D2D model selection for load balancing. Moreover, the offloaded users and D2D receivers (RXs) would suffer strong interference from BSs, especially from high-power BSs. Evidently, a good association scheme should integrate with interference mitigation. Thus, we first propose an effective resource partitioning strategy that can mitigate the interference received by offloaded users from high-power BSs and the one received by D2D RXs from BSs. Based on this, we then design a user association scheme for load balancing, which jointly considers user association and D2D model selection to maximize network-wide utility. Considering that the formulated problem is in a nonlinear and mixted-integer form and hard to tackle, we adopt a dual decomposition method to develop an efficient distributed algorithm. Simulation results show that the proposed scheme provides a load balancing gain and a resource partitioning gain.
\end{abstract}


%
\IEEEpeerreviewmaketitle

\section{Introduction}
%
%
%
%
\IEEEPARstart{H}{e}terogeneous cellular networks (HCNs) have been widely regarded as a promising solution to improving area's spectral efficiency, alleviating traffic congestions in hot-spots, and thus enhancing the end-user experience \cite{1,2,3,4}. However, due to the limited backhaul connections between low-power base stations (BSs), some offloading techniques cannot fully balance the loads among different BSs. It is highly possible that some BSs are severely congested while adjacent BSs are very lightly loaded. To further alleviate congestions and increase system throughput, HCNs with device-to-device (D2D) communications have been a good option in recent years \cite{5,6,7}.
\par
The D2D communication directly takes place between two closely located users. As a kind of proximity communication, it has attracted more and more attention due to its advantages such as offloading traffic, low power consumption and latency, and high data rate and spectral efficiency \cite{8,9}. Moreover, the D2D communications can also enhance the physical layer security \cite{10}. Thus, the D2D communications have been advocated in various wireless networks. Although D2D communications can achieve many advantages, the D2D pairs may suffer severe interference from BSs, especially from high-power BSs. To fully exploit the potential of D2D communications, we need to consider some interference mitigation techniques such as power control and resource partitioning \cite{11}.
\par
Since various BSs coexist, the user association for HCNs is a challenging topic \cite{12}. When the conventional signal strength-based association is applied to HCNs, the obtained load distribution is very imbalanced since most users are associated with high-power BSs. Thus, some association approaches that are well applied in traditional cellular networks may not be appropriate for HCNs. Moreover, when D2D communication techniques are incorporated into HCNs, the user association problem becomes more complicated. To make full use of novel network framework, we are required to design an association scheme with offloading capability. Next, we will focus on some offloading strategies for HCNs and D2D-enabled cellular networks.
\subsection{Related work}
To accommodate new characteristics of HCNs, many efforts in the literature toward load balancing for HCNs have been made. As a most frequently utilized method, the biasing method (cell range expansion) gives low-power BSs an offset to attract more users for them. In \cite{13}, authors give closed-form expressions of the downlink data rate and signal-interference-plus-noise ratio (SINR) distribution under a biasing method, but cannot achieve a closed-form expression of an optimal offset. Since this association approach only focuses on the load balancing for HCNs and doesn't consider any interference mitigation measures, the offloaded users may receive the strong interference from high-power BSs. In order to avoid this interference, Singh et al. \cite{14} propose a resource partitioning scheme so that the system throughput is greatly improved. Moreover, Jo et al. \cite{15} also develop a tractable framework for the downlink SINR and rate analysis under a biasing method. Although the biasing method is simple for load balancing in HCNs, the closed-form expression of an optimal offset is not provided in existing works.
\par
Besides the biasing method, there are other offloading schemes for HCNs. In \cite{16}, authors perform the user association to maximize a sum-utility that relates to long-term rates, and develop an efficient distributed user association algorithm. Considering the cross-tier interference, Shen et al. \cite{17} design a user association scheme with power control or beamforming based on \cite{16}, and put forward an improved algorithm. Unlike \cite{16}, \cite{18} studies the interplay of user association and resource partitioning, and tries to design an association algorithm with guaranteeing the upper bounds of system performance. Recently, Cho et al. \cite{19} adopt repulsive cell activation in the interfering daughtercell network to balance cell loads, but this method appears to be high complicated.
 \par
Due to the limited backhaul connections between low-power BSs, most of aforementioned schemes may not achieve efficient load balancing. In other words, some offloading techniques cannot fully balance the loads among different BSs. To fully exploit the potential of D2D communications and thus alleviate network congestions, the user association for load balancing in D2D-enabled HCNs has been investigated in \cite{5,6}. In \cite{5}, authors propose an online offloading scheme that doesn't consider interference mitigation. Moreover, the D2D pairs in \cite{5} just play the role of relay and cannot support direct data communication. Unlike \cite{5}, the D2D receivers (RXs) in \cite{6} can directly communicate with D2D transmitters (TXs) or other BSs. Authors in \cite{6} jointly consider user association, D2D model selection and power control for uplink D2D-enabled HCNs. Note that the D2D model selection represents that a D2D RX selects some D2D TX or BS to connect. So far, few existing efforts jointly consider user association and D2D model selection for load balancing in downlink D2D-enabled HCNs. It is necessary to design an efficient user association algorithm, which balances the loads among BSs and fully utilizes D2D communications.
\subsection{Contributions and organization}
In this paper, we propose a load balancing scheme for downlink D2D-enabled HCNs, which jointly considers the user association for cellular users and the D2D model selection for potential D2D pairs. Moreover, in order to mitigate the interference received by offloaded users from high-power BSs and the one received by D2D RXs from BSs, we design a resource partitioning scheme. Specially, the whole frequency band is cut into three subbands including subbands 1, 2 and 3, where the high-power BSs monopolize subband 1, the D2D TXs can also monopolize subband 3 and the low-power BSs can utilize subbands 1 and 2. In this way, some offloaded users can be associated with the subband 2 of low-power BSs to avoid the strong interference from high-power BSs, some D2D RXs can be associated with the subband 3 of D2D TXs to avoid the interference from all BSs. At last, the load balancing scheme is formulated as a network-wide utility maximization problem. Considering that the formulated problem is in a nonlinear and mixted-integer form and hard to tackle, we utilize a dual decomposition method to develop an efficient distributed algorithm.
\par
The rest of this paper is organized as follows. In Section 2, we describe our system model including network model and resource partitioning model. In Section 3, we formulate the user association problem. In Section 4, we design a distributed algorithm using dual decomposition. In Section 5, we discuss our simulations on a load balancing gain and a resource partitioning gain. In Section 6, we present further discussions and conclusions.
\begin{figure}[!t]
\centering
\centerline{\includegraphics[width=4in]{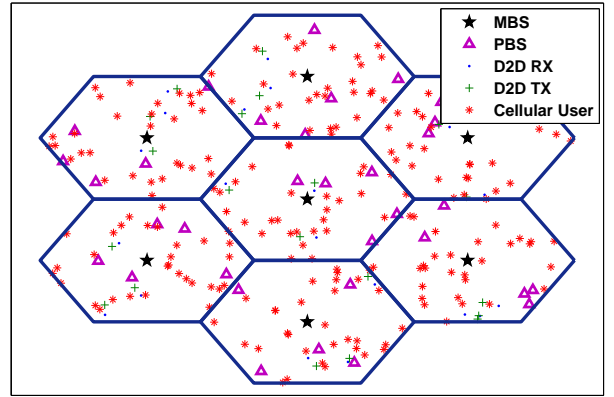}}
\caption{Illustration of D2D-enabled HCNs. In the HCNs, MBSs are deployed into a conventional cellular network, potential D2D pairs, PBSs and cellular users are uniformly and independently scattered into each macrocell.}
\label{fig1}
\end{figure}
\section{System model}
In this paper, we consider a D2D-enabled HCN that is the conventional macrocellular network overlaid with pico BSs (PBSs) and potential D2D pairs. This network deployment is illustrated in Fig. \ref{fig1}, where MBS is the macro BS, and each D2D pair contains D2D RX and D2D TX. In general, cellular users, D2D TXs and D2D RXs are called users, PBSs and MBSs are called BSs. Moreover, cellular users, D2D TXs and D2D RXs are also called receivers, PBSs, MBSs and D2D TXs are also called transmitters. Note that the D2D TXs are the receivers of BSs and they are also the transmitters of the corresponding D2D RXs.
\par
To proceed, we need to make the following assumption.
\par
\noindent
\textbf{Assumption 1.} Each BS equally allocates power to all subbands being employed.
\par
\noindent
\emph{Remark: }This assumption has been widely used for downlink resource allocation due to its implementation simplicity and analytical tractability. Moreover, equal power allocation can achieve near-optimal solutions in many cases, especially at high SINR regime \cite{20,21,22}.
\par
\begin{figure}[!t]
\centering
\centerline{\includegraphics[width=3.5in]{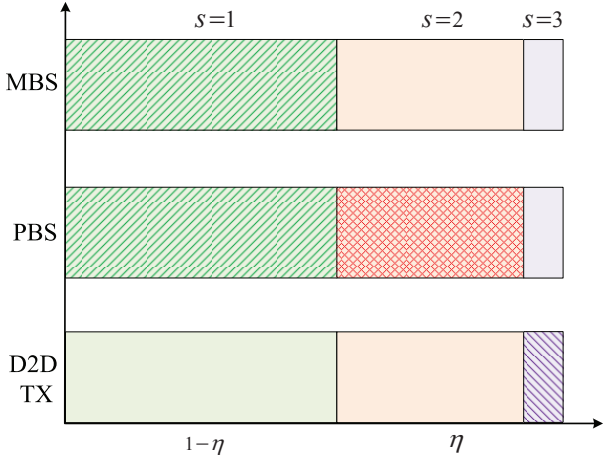}}
\caption{Resource partitioning.}
\label{fig2}
\end{figure}

To reduce the interference received by offloaded users and D2D RXs from MBSs, we introduce a resource partitioning scheme. Specially, the whole frequency band is cut into three subbands including subband 1 ($s=1$), subband 2 ($s=2$) and subband 3 ($s=3$). As illustrated in Fig. \ref{fig2}, MBSs just utilize subband 1, D2D TXs just use subband 3, and PBSs can utilize subbands 1 and 2. Note that the bandwidths of subbands 1, 2 and 3 are $\left(1-\eta\right){{\text{W}}_1}$, $\eta {{\text{W}}_1}$ and ${{\text{W}}_2}$ respectively, where ${{\text{W}}_1}=\text{W}-{{\text{W}}_2}$, ${{\text{W}}_2}={{\text{W}}_{\text{prb}}}$, $\text{W}$ is the system bandwidth and ${{\text{W}}_{\text{prb}}}$ represents the bandwidth of one PRB (physic resource block). According to the descriptions of Long Term Evolution (LTE)\cite{23}, adjacent twelve subcarriers are grouped into one PRB with 180KHz, which is the smallest unit that can be allocated to each user. Considering that one D2D TX just serves one user (the corresponding D2D RX), one PRB may be sufficient for D2D TX. In this way, some offloaded users can be associated with the subband 1 of PBSs to avoid the strong interference from MBSs, some D2D RXs can be associated with the subband 3 of D2D TXs to avoid the interference from all BSs. Under the resource partitioning scheme, all users may be associated with the subband 1 of MBSs or some subband of PBSs. Moreover, any D2D RX may also be associated with the subband 3 of its corresponding D2D TX.
\par
Now, we let the set of MBSs be ${{\mathcal{N}}_{m}}$, let the set of PBSs be ${{\mathcal{N}}_{p}}$, let the set of cellular users be ${{\mathcal{K}}_{c}}$, let the set of D2D TXs be ${{\mathcal{K}}_{t}}$ and denote the set of D2D RXs as ${{\mathcal{K}}_{r}}$, where $\mathcal{N}={{\mathcal{N}}_{m}}\cup {{\mathcal{N}}_{p}}\cup {{\mathcal{K}}_{t}}$, ${{\mathcal{N}}_{mp}}={{\mathcal{N}}_{m}}\cup {{\mathcal{N}}_{p}}$ and $\mathcal{K}={{\mathcal{K}}_{c}}\cup {{\mathcal{K}}_{t}}\cup {{\mathcal{K}}_{r}}$. Meanwhile, we write the set of subbands as $S=\left\{ 1,2,3 \right\}$, and write the cardinalities of sets  $\mathcal{N}$, $\mathcal{N}_{mp}$, $\mathcal{N}_{m}$, $\mathcal{N}_{p}$, $\mathcal{S}$ and $\mathcal{K}$ as $N=\left| \mathcal{N} \right|$, $N_{mp}=\left| \mathcal{N}_{mp} \right|$, $N_{m}=\left| \mathcal{N}_{m} \right|$, $N_{p}=\left| \mathcal{N}_{p} \right|$, $S=\left| \mathcal{S} \right|$ and $K=\left| \mathcal{K} \right|$ respectively. Then, the SINR received by user $k\in {{\mathcal{K}}}$ from BS $n\in {{\mathcal{N}}_{mp}}$ on subband 1 can be expressed as
\begin{equation}\label{eq1}
{{\text{SINR}}_{n1k}}=\frac{{{p}_{n1}}{{g}_{n1k}}}{{{\sum }_{j\in {{\mathcal{N}}_{mp}}\backslash \left\{ n \right\}}}{{p}_{j1}}{{g}_{j1k}}+\left( 1-\eta \right) {{\text{W}}_1}{{\text{N}}_{0}}},
\end{equation}
\par
\noindent
and the SINR received by user ${k\in \mathcal{K}}$ from BS ${\ n\in {{\mathcal{N}}_{p}}}$ on subband 2 can be expressed as
\begin{equation}\label{eq2}
{{\text{SINR}}_{n2k}}=\frac{{{p}_{n2}}{{g}_{n2k}}}{{{\sum }_{j\in {{\mathcal{N}}_{p}}\backslash \left\{ n \right\}}}{{p}_{j2}}{{g}_{j2k}}+\eta {{\text{W}}_1}{{\text{N}}_{0}}},
\end{equation}
\par
\noindent
and the SINR received by D2D RX $k\in {{\mathcal{K}}_{r}}$ from the corresponding D2D TX ${{n}_{k}}$ on subband 3 can be expressed as
\begin{equation}\label{eq3}
{{\text{SINR}}_{n_{k}3k}}=\frac{{{p}_{n_{k}3}}{{g}_{n_{k}3k}}}{{{\sum }_{j\in {{\mathcal{N}}_{t}}\backslash \left\{ n_{k} \right\}}}{{p}_{j3}}{{g}_{j3k}}+\eta {{\text{W}}_2}{{\text{N}}_{0}}},
\end{equation}
where ${{p}_{ns}}$ represents the transmit power of transmitter $n$ on subband $s$; ${{g}_{nsk}}$ denotes the channel gain between receiver $k$ and transmitter $n$ on subband $s$; ${{\text{N}}_{0}}$ is the noise power spectral density. Since the D2D TXs also need to be associated BSs, the D2D TX $k\in {{\mathcal{K}}_{t}}$ can be regarded as the receiver of BSs, but it can also be seen as the transmitter of the corresponding D2D RX.
\par
Then, the achievable rate [in bps] received by user $k\in {{\mathcal{K}}}$ from BS $n\in {{\mathcal{N}}_{mp}}$ on subband 1 can be expressed as
\begin{equation}\label{eq4}
{{r}_{n1k}}=\left( 1-\eta  \right){{\text{W}}_1}{{\log }_{2}}\left( 1+{{\text{SINR}}_{n1k}} \right),
\end{equation}
\par
\noindent
and the achievable rate [in bps] received by user ${k\in \mathcal{K}}$ from BS ${\ n\in {{\mathcal{N}}_{p}}}$ on subband 2 can be expressed as
\begin{equation}\label{eq5}
{{r}_{n2k}}=\eta {{\text{W}}_1}{{\log }_{2}}\left( 1+{{\text{SINR}}_{n2k}} \right),
\end{equation}
\par
\noindent
and the achievable rate [in bps] received by D2D RX $k\in {{\mathcal{K}}_{r}}$ from the corresponding D2D TX ${{n}_{k}}$ on subband 3 can be expressed as
\begin{equation}\label{eq6}
{{r}_{n_{k}3k}}={{\text{W}}_2}{{\log }_{2}}\left( 1+{{\text{SINR}}_{n_{k}3k}} \right),
\end{equation}
\par
Moreover, since some subbands of some transmitters cannot be utilized by receivers, the achievable rates can be set to 0 in these cases. In order to meet the demand of algorithm design, i.e., avoid the case $\log \left( 0 \right)$,  we add a very small constant $\vartheta $ to the achievable rate. Thus, we have ${{r}_{nsk}}={{r}_{nsk}}+\vartheta $, e.g., $\vartheta=10^{-20}$.
\par
To proceed, we need to give the following definitions.
\par
\noindent
\textbf{Definition 1.} The effective load of transmitter $n$ on subband $s$ is ${{y}_{ns}}=\sum\nolimits_{k\in \mathcal{K}}{{{x}_{nsk}}}$, where ${{x}_{nsk}}$ represents the association indicator, i.e., ${{x}_{nsk}}=1$ when receiver $k$ is associated with subband $s$ of transmitter $n$, 0 otherwise.
\par
\noindent
\textbf{Definition 2.} If the load of transmitter $n$ on the subband $s$ is ${{y}_{ns}}$, the effective rate of receiver $k$ who is associated with subband $s$ of transmitter $n$ is given by ${{R}_{nsk}}={{r}_{nsk}}/{{y}_{ns}}$.
\section{Problem formulation}
In this section, we formulate the joint user association and resource partitioning problem to maximize network-wide utility. Specially, this problem is formulated as
\begin{equation}\label{eq7}
\begin{split}
  \underset{\boldsymbol{x}}{\mathop{\max }}\,&\ F\left( \boldsymbol{x} \right)=\sum\limits_{n\in \mathcal{N}}{\sum\limits_{s\in \mathcal{S}}{\sum\limits_{k\in \mathcal{K}}{{{x}_{nsk}}{{U}_{nsk}}\left( {{R}_{nsk}} \right)}}} \\
 \text{s.t. }&\sum\limits_{n\in \mathcal{N}}{\sum\limits_{s\in \mathcal{S}}{{{x}_{nsk}}}}=1,\ \forall k\in \mathcal{K}, \\
 & {{x}_{nsk}}\in \left\{ 0,1 \right\},\ \forall n\in \mathcal{N},\forall s\in \mathcal{S},\forall k\in \mathcal{K}, \\
\end{split}
\end{equation}
where $\boldsymbol{x}=\left\{ {{x}_{nsk}},n\in \mathcal{N},s\in \mathcal{S},k\in \mathcal{K} \right\}$; ${{U}_{nsk}}$ denotes the utility received by receiver $k$ from transmitter $n$ on subband $s$; the first constraint means that a receiver can be only connected to one transmitter on some subband.
\par
\noindent
\textbf{Proposition 1. } Under resource partitioning, the equivalent form of the problem \eqref{eq7} is given by
\begin{equation}\label{eq8}
\begin{split}
  \underset{\boldsymbol{x}}{\mathop{\max }}\,&\ {{G}}\left( \boldsymbol{x} \right) \\
 \text{s.t. }&\sum\limits_{n\in \mathcal{N}}{\sum\limits_{s\in \mathcal{S}}{{{x}_{nsk}}}}=1,\ \forall k\in \mathcal{K}, \\
 & {{x}_{nsk}}\in \left\{ 0,1 \right\},\ \forall n\in \mathcal{N},\forall s\in \mathcal{S},\forall k\in \mathcal{K}, \\
\end{split}
\end{equation}
where ${{c}_{nsk}}=\log {{r}_{nsk}}$ and
\begin{equation}\label{eq9}
\begin{split}
  {{G}}\left( \boldsymbol{x} \right)=&\sum\limits_{n\in {{\mathcal{N}}_{mp}}}{\sum\limits_{k\in {{\mathcal{K}}}}{{{x}_{n1k}}\left( {{c}_{n1k}}-\log \sum\limits_{j\in {{\mathcal{K}}}}{{{x}_{n1j}}} \right)}} \\
 & +\sum\limits_{n\in {{\mathcal{N}}_{p}}}{\sum\limits_{k\in {{\mathcal{K}}}}{{{x}_{n2k}}\left( {{c}_{n2k}}-\log \sum\limits_{j\in {{\mathcal{K}}}}{{{x}_{n2j}}} \right)}} \\
 & +\sum\limits_{k\in {{\mathcal{K}}_{r}}}{{{x}_{{{n}_{k}}3k}}\left( {{c}_{{{n}_{k}}3k}}-\log {{x}_{{{n}_{k}}3k}} \right)}, \\
\end{split}
\end{equation}
\par
\emph{Proof}: According to the definition of achievable rate, we know that the achievable rates of MBSs on subband 1, the ones of PBSs on subbands 1 and 2, and the ones received by any D2D RX $k\in {{\mathcal{K}}_{r}}$ from the corresponding D2D TX ${{n}_{k}}$ on subband 3 are larger than $\vartheta$. In other cases, the achievable rates equal to $\vartheta$. When the achievable rates are $\vartheta$, these terms in the objective function $F\left( \boldsymbol{x} \right)$ of the problem \eqref{eq7} don't need to be considered. In fact, the achievable rates with $\vartheta$ mean that some subbands of transmitters are unavailable. Thus, we have
\begin{equation}\label{eq10}
\begin{split}
  F\left( \boldsymbol{x} \right)\equiv & G\left( \boldsymbol{x} \right)\\
 =&\sum\limits_{n\in {{\mathcal{N}}_{mp}}}{\sum\limits_{k\in {{\mathcal{K}}}}{{{x}_{n1k}}{{U}_{n1k}}\left( {{R}_{n1k}} \right)}}\\
 &+\sum\limits_{n\in {{\mathcal{N}}_{p}}}{\sum\limits_{k\in \mathcal{K}}{{{x}_{n2k}}{{U}_{n2k}}\left( {{R}_{n2k}} \right)}} \\
 &+\sum\limits_{k\in {{\mathcal{K}}_{r}}}{{{x}_{{{n}_{k}}3k}}{{U}_{{{n}_{k}}3k}}\left( {{R}_{{{n}_{k}}3k}} \right)}. \\
\end{split}
\end{equation}
\par
To guarantee the user fairness, we introduce a logarithmic function as the mentioned utility function of problem \eqref{eq7}. Then, we have the expression \eqref{eq9}.
\par
Seen from the objective function \eqref{eq9}, the receiver needs to trade off the load and achievable rate when it selects the subband 1 of transmitter (BS) $n\in {{\mathcal{N}}_{mp}}$ or the subband 2 of BS $n\in {{\mathcal{N}}_{p}}$. In other words, the proposed scheme is not the maximal rate (Max-Rate) association but the association that owns an offloading capability.
\par
According to the definition of effective load, we can further convert problem \eqref{eq8} into
\begin{equation}\label{eq11}
\begin{split}
  \underset{\boldsymbol{x},\boldsymbol{y}}{\mathop{\max }}\,&\ H\left( \boldsymbol{x},\boldsymbol{y} \right) \\
\text{s.t. }&\sum\limits_{n\in \mathcal{N}}{\sum\limits_{s\in \mathcal{S}}{{{x}_{nsk}}}}=1,\ \forall k\in \mathcal{K}, \\
 &\sum\limits_{k\in {{\mathcal{K}}}}{{{x}_{n1k}}}={{y}_{n1}},\ \forall n\in {{\mathcal{N}}_{mp}}, \\
 &\sum\limits_{k\in \mathcal{K}}{{{x}_{n2k}}}={{y}_{n2}},\ \forall n\in {{\mathcal{N}}_{p}}, \\
 &{{x}_{nsk}}\in \left\{ 0,1 \right\},\ \forall n\in \mathcal{N},\forall s\in \mathcal{S},\forall k\in \mathcal{K}, \\
\end{split}
\end{equation}
where $\boldsymbol{y}=\left\{ {{y}_{ns}},\ \ \forall n\in \mathcal{N},\forall s\in \mathcal{S}\right\}$ and
\begin{equation}\label{eq12}
\begin{split}
  H\left( \boldsymbol{x},\boldsymbol{y} \right)=&\sum\limits_{n\in {{\mathcal{N}}_{mp}}}{\sum\limits_{k\in {{\mathcal{K}}}}{{{x}_{n1k}}{{c}_{n1k}}}}-\sum\limits_{n\in {{\mathcal{N}}_{mp}}}{{{y}_{n1}}\log {{y}_{n1}}} \\
 &+\sum\limits_{n\in {{\mathcal{N}}_{p}}}{\sum\limits_{k\in {{\mathcal{K}}}}{{{x}_{n2k}} {{c}_{n2k}}}-\sum\limits_{n\in {{\mathcal{N}}_{p}}}{{{y}_{n2}}\log {{y}_{n2}}}} \\
 &+\sum\limits_{k\in {{\mathcal{K}}_{r}}}{{{x}_{{{n}_{k}}3k}}\left( {{c}_{{{n}_{k}}3k}}-\log {{x}_{{{n}_{k}}3k}} \right)}. \\
\end{split}
\end{equation}
\
\section{Association algorithm}
To find the global optimal solutions of the problem \eqref{eq11}, global network information should be collected. That means a centralized controller is required to perform user association and coordination. Considering this case, we design a distributed algorithm for the proposed user association problem to achieve suboptimal solutions, which doesn't require any coordination among BSs.
\par
As a general method to solve an optimization problem, the dual decomposition has attracted increasing attention in the literature since it can simplify a highly complex and large-scale problem. Specially, this method breaks the original optimization problem up into smaller subproblems and separately solves these smaller ones in a distributed manner. So far, the dual decomposition method has been widely used in user association \cite{16,17}, simultaneous routing and resource allocation  \cite{24}, sum capacity maximization \cite{25} and distributed control \cite{26}. In this section, we design a highly efficient distributed algorithm by dual decomposition. Then, users and BSs can separately solve their two subproblems into which the dual problem is decoupled.
\par
Considering the coupling constraints in the problem \eqref{eq11}, i.e., the second constraint and the third constraint, we introduce the Lagrange multipliers to relax them. For any BS $n\in {{\mathcal{N}}_{mp}}$, we introduce ${{\mu }_{n1}}$ for the corresponding (second) constraint; For any BS $n\in {{\mathcal{N}}_{p}}$, we introduce ${{\mu }_{n2}}$ for the corresponding (third) constraint. Thus, the Lagrange function with respect to these constraints is
\begin{equation}\label{eq13}
\begin{split}
  \mathcal{L}\left( \boldsymbol{x},\boldsymbol{y},\boldsymbol{u} \right)=& \sum\limits_{n\in {{\mathcal{N}}_{mp}}}{\sum\limits_{k\in {{\mathcal{K}}}}{{{x}_{n1k}}{{c}_{n1k}}}}-\sum\limits_{n\in {{\mathcal{N}}_{mp}}}{{{y}_{n1}}\log {{y}_{n1}}} \\
 & +\sum\limits_{n\in {{\mathcal{N}}_{p}}}{\sum\limits_{k\in \mathcal{K}}{{{x}_{n2k}}{{c}_{n2k}}}}-\sum\limits_{n\in {{\mathcal{N}}_{p}}}{{{y}_{n2}}\log {{y}_{n2}}} \\
 & +\sum\limits_{k\in {{\mathcal{K}}_{r}}}{{{x}_{{{n}_{k}}3k}}\left( {{c}_{{{n}_{k}}3k}}-\log {{x}_{{{n}_{k}}3k}} \right)} \\
 & +\sum\limits_{n\in {{\mathcal{N}}_{mp}}}{{{\mu }_{n1}}\left( {{y}_{n1}}-\sum\limits_{k\in {{\mathcal{K}}}}{{{x}_{n1k}}} \right)} \\
 & +\sum\limits_{n\in {{\mathcal{N}}_{p}}}{{{\mu }_{n2}}\left( {{y}_{n2}}-\sum\limits_{k\in \mathcal{K}}{{{x}_{n2k}}} \right)}. \\
\end{split}
\end{equation}
\par
Then, the dual function can be written as
\begin{equation}\label{eq14}
I\left( \boldsymbol{\mu } \right)=\left\{ \begin{split}
  &\underset{\boldsymbol{x},\boldsymbol{y}}{\mathop{\max }}\,\ \mathcal{L}\left( \boldsymbol{x},\boldsymbol{y},\boldsymbol{\mu } \right) \\
&\text{s.t. }\sum\limits_{n\in \mathcal{N}}{\sum\limits_{s\in \mathcal{S}}{{{x}_{nsk}}}}=1,\ \forall k\in \mathcal{K}, \\
 & {{x}_{nsk}}\in \left\{ 0,1 \right\},\ \forall n\in \mathcal{N},\forall s\in \mathcal{S},\forall k\in \mathcal{K}, \\
\end{split} \right.
\end{equation}
and the dual problem of \eqref{eq11} is given by
\begin{equation}\label{eq15}
\underset{\boldsymbol{\mu }}{\mathop{\min }}\,I\left( \boldsymbol{\mu } \right).
\end{equation}
\par
Considering the problem \eqref{eq15} is not coupling with respect to $\boldsymbol{x}$ and $\boldsymbol{y}$, we can separately obtain the primal optimal solutions. Thus, the problem \eqref{eq15} can be decomposed into
\begin{equation}\label{eq16}
{{I}_{1}}\left( \boldsymbol{\mu } \right)=\left\{ \begin{split}
  & \underset{\boldsymbol{x}}{\mathop{\max }}\,\ {{\mathcal{L}}_{1}}\left( \boldsymbol{x},\boldsymbol{\mu } \right) \\
 &\text{s.t. }\sum\limits_{n\in \mathcal{N}}{\sum\limits_{s\in \mathcal{S}}{{{x}_{nsk}}}}=1,\ \forall k\in \mathcal{K}, \\
 &{{x}_{nsk}}\in \left\{ 0,1 \right\},\ \forall n\in \mathcal{N},\forall s\in \mathcal{S},\forall k\in \mathcal{K}, \\
\end{split} \right.
\end{equation}
\par
\noindent
and
\par
\noindent
\begin{equation}\label{eq17}
{{I}_{2}}\left( \boldsymbol{\mu } \right)=\underset{\boldsymbol{y}}{\mathop{\max }}\,\ {{\mathcal{L}}_{2}}\left( \boldsymbol{y},\boldsymbol{\mu } \right),
\end{equation}
where
\begin{equation}\label{eq18}
\begin{split}
  {{\mathcal{L}}_{1}}\left( \boldsymbol{x},\boldsymbol{u} \right)=& \sum\limits_{n\in {{\mathcal{N}}_{mp}}}{\sum\limits_{k\in {{\mathcal{K}}}}{{{x}_{n1k}}\left( {{c}_{n1k}}-{{\mu }_{n1}} \right)}}\\
  &+\sum\limits_{n\in {{\mathcal{N}}_{p}}}{\sum\limits_{k\in \mathcal{K}}{{{x}_{n2k}}\left( {{c}_{n2k}}-{{\mu }_{n2}} \right)}} \\
 &+\sum\limits_{k\in {{\mathcal{K}}_{r}}}{{{x}_{{{n}_{k}}3k}}\left( {{c}_{{{n}_{k}}3k}}-\log {{x}_{{{n}_{k}}3k}} \right)}, \\
\end{split}
\end{equation}
and
\begin{equation}\label{eq19}
\begin{split}
{{\mathcal{L}}_{2}}\left( \boldsymbol{y},\boldsymbol{u} \right)=&\sum\limits_{n\in {{\mathcal{N}}_{mp}}}{{{y}_{n1}}\left( {{\mu }_{n1}}-\log {{y}_{n1}} \right)}\\
&+\sum\limits_{n\in {{\mathcal{N}}_{p}}}{{{y}_{n2}}\left( {{\mu }_{n2}}-\log {{y}_{n2}} \right)}.
\end{split}
\end{equation}
\par
When the dual optimal ${{\boldsymbol{\mu} }^{*}}$ is given, the optimal solutions of \eqref{eq11} can be obtained by separately solving its two subproblems.
\par
Now, we solve the outer problem \eqref{eq15} by a gradient projection method \cite{27}. For any BS $n\in {{\mathcal{N}}_{mp}}$, we search the optimal ${{\mu }_{n1}}$ in the direction of negative gradient, i.e.,$-\nabla G\left( {{\mu }_{n1}}  \right)$. Similarly, for any BS $n\in {{\mathcal{N}}_{p}}$, we search the optimal ${{\mu }_{n2}}$ in the direction of negative gradient, i.e.,$-\nabla G\left( {{\mu }_{n2}}  \right)$. To obtain these gradients, we need to solve subproblem \eqref{eq16} and \eqref{eq17}.
\par
According the condition of resource utilization, we can give the expanded form of the expression \eqref{eq18}. Then, we have
\begin{equation}\label{eq20}
\begin{split}
  {{\mathcal{L}}_{1}}\left( \boldsymbol{x},\boldsymbol{u} \right)=&\sum\limits_{n\in {{\mathcal{N}}_{mp}}}{\sum\limits_{k\in {{\mathcal{K}}_{c}}}{{{x}_{n1k}}\left( {{c}_{n1k}}-{{\mu }_{n1}} \right)}} \\
 &+\sum\limits_{n\in {{\mathcal{N}}_{mp}}}{\sum\limits_{k\in {{\mathcal{K}}_{t}}}{{{x}_{n1k}}\left( {{c}_{n1k}}-{{\mu }_{n1}} \right)}} \\
  &+\sum\limits_{n\in {{\mathcal{N}}_{mp}}}{\sum\limits_{k\in {{\mathcal{K}}_{r}}}{{{x}_{n1k}}\left( {{c}_{n1k}}-{{\mu }_{n1}} \right)}} \\
 &+\sum\limits_{n\in {{\mathcal{N}}_{p}}}{\sum\limits_{k\in {{\mathcal{K}}_{c}}}{{{x}_{n2k}}\left( {{c}_{n2k}}-{{\mu }_{n2}} \right)}} \\
 &+\sum\limits_{n\in {{\mathcal{N}}_{p}}}{\sum\limits_{k\in {{\mathcal{K}}_{t}}}{{{x}_{n2k}}\left( {{c}_{n2k}}-{{\mu }_{n2}} \right)}} \\
 &+\sum\limits_{n\in {{\mathcal{N}}_{p}}}{\sum\limits_{k\in {{\mathcal{K}}_{r}}}{{{x}_{n2k}}\left( {{c}_{n2k}}-{{\mu }_{n2}} \right)}} \\
 &+\sum\limits_{k\in {{\mathcal{K}}_{r}}}{{{x}_{{{n}_{k}}3k}}\left( {{c}_{{{n}_{k}}3k}}-\log {{x}_{{{n}_{k}}3k}} \right)}. \\
\end{split}
\end{equation}
Moreover, the equation \eqref{eq20} is equivalent to
\begin{equation}\label{eq21}
\begin{split}
  {{\mathcal{L}}_{1}}\left( \mathbf{x},\mathbf{u} \right)=&\sum\limits_{k\in {{\mathcal{K}}_{c}}}{\left\{ \sum\limits_{n\in {{\mathcal{N}}_{mp}}}{{{x}_{n1k}}\left( \log {{r}_{n1k}}-{{\mu }_{n1}} \right)} \right.} \\
 & \left. \ \ \ \ \ \ \ \ \ \ \ \  +\sum\limits_{n\in {{\mathcal{N}}_{p}}}{{{x}_{n2k}}\left( \log {{r}_{n2k}}-{{\mu }_{n2}} \right)} \right\} \\
 & +\sum\limits_{k\in {{\mathcal{K}}_{r}}}{\left\{ \sum\limits_{n\in {{\mathcal{N}}_{mp}}}{{{x}_{n1k}}\left( \log {{r}_{n1k}}-{{\mu }_{n1}} \right)} \right.} \\
 & \ \ \ \ \ \ \ \ \ \ \ \ +{{x}_{{{n}_{k}}3k}}\left( \log {{r}_{{{n}_{k}}3k}}-\log {{x}_{{{n}_{k}}3k}} \right) \\
 & \left. \ \ \ \ \ \ \ \ \ \ \ \ +\sum\limits_{n\in {{\mathcal{N}}_{p}}}{{{x}_{n2k}}\left( \log {{r}_{n2k}}-{{\mu }_{n2}} \right)} \right\} \\
 & +\sum\limits_{k\in {{\mathcal{K}}_{t}}}{\left\{ \sum\limits_{n\in {{\mathcal{N}}_{mp}}}{{{x}_{n1k}}\left( \log {{r}_{n1k}}-{{\mu }_{n1}} \right)} \right.} \\
 & \left. \ \ \ \ \ \ \ \ \ \ \ \ +\sum\limits_{n\in {{\mathcal{N}}_{p}}}{{{x}_{n2k}}\left( \log {{r}_{n2k}}-{{\mu }_{n2}} \right)} \right\} \\
\end{split}
\end{equation}
\par
According the form of the subproblem \eqref{eq16}, we can easily develop its algorithm whose detailed procedure can be found in Algorithm 1. Since this algorithm is performed by users including cellular users, D2D RXs and D2D TXs, it can be regarded as the algorithm on user's side. In the steps 6-10 of Algorithm 1, cellular user $k\in {{\mathcal{K}}_{c}}$ selects the subband 1 of some BS or the subband 2 of some PBS to maximize its utility, i.e., achieve the maximal utility ${{c}_{n^{*}s^{*}k}}-{{\mu }_{n^{*}s^{*}}}$. In other words, user $k$ selects the subband $s^{*}$ of BS $n^{*}$ if ${{c}_{n^{*}s^{*}k}}-{{\mu }_{n^{*}s^{*}}}$ is the maximal value among possible associations. As mentioned in previous section, the D2D RX $k\in {{\mathcal{K}}_{r}}$ can be associated with the subband 1 of some BS or the subband 2 of some PBS or the corresponding D2D TX $n_{k}$. When the D2D TX is not utilized by the corresponding D2D RX, the term ${{x}_{{{n}_{k}}3k}}\left( {{c}_{{{n}_{k}}3k}}-\log {{x}_{{{n}_{k}}3k}} \right)$ in the association object can be neglected. However, when the D2D TX is selected by the corresponding D2D RX, the term ${{x}_{{{n}_{k}}3k}}\left( {{c}_{{{n}_{k}}3k}}-\log {{x}_{{{n}_{k}}3k}} \right)$ can be simplified into $ {{c}_{{{n}_{k}}3k}} $. Thus, in the steps 11-18 of Algorithm 1, the D2D RX $k\in {{\mathcal{K}}_{r}}$ first selects the subband 1 of some BS or the subband 2 of some PBS to achieve the maximal utility ${{c}_{n^{*}s^{*}k}}-{{\mu }_{n^{*}s^{*}}}$, then selects the subband 3 of the corresponding D2D TX ${{n}_{k}}$ if ${{c}_{n^{*}s^{*}k}}-{{\mu }_{n^{*}s^{*}}}<{{c}_{n_{k}3k}}$. Similar to the cellular users, the D2D TX $k\in {{\mathcal{K}}_{t}}$ perform user association in the steps 19-23 of Algorithm 1.
\begin{table}[h]
\centering
\begin{tabular}{ll}
\toprule[1pt]
\textbf{Algorithm 1 at User Terminal $k$} \\ \midrule[0.5pt]
1: \ \textbf{If } $t=0$ \\
2: \ \ \ \ Estimate $\boldsymbol{c}$ using pilots signals from all transmitters. \\
3: \ \textbf{Else } \\
4: \ \ \ \ Receive the information ${{\mu }_{n1}}$ broadcasted by all BSs.\\
5: \ \ \ \ Receive the information ${{\mu }_{n2}}$ broadcasted by all PBSs.\\
6: \ \ \ \textbf{If} $k\in {{\mathcal{K}}_{c}}$\\
7: \ \ \ \  \ \ \ Select the subband 1 of some BS or the subband 2 of some\\
8: \ \ \ \  \ \ \ PBS to achieve the maximal utility: ${{c}_{n^{*}s^{*}k}}-{{\mu }_{n^{*}s^{*}}}$.\\
9: \ \ \ \  \ \ \ $x_{n^{*}s^{*}k}=1$. \\
10: \ \ \ \textbf{EndIf}\\
11: \ \ \ \textbf{If} $k\in {{\mathcal{K}}_{r}}$\\
12: \ \ \ \  \ \ \ Select the subband 1 of some BS or the subband 2 of some\\
13: \ \ \ \  \ \ \ PBS to achieve the maximal utility: ${{c}_{n^{*}s^{*}k}}-{{\mu }_{n^{*}s^{*}}}$.\\
14: \ \ \ \  \ \ \ \ \textbf{If} ${{c}_{n^{*}s^{*}k}}-{{\mu }_{n^{*}s^{*}}}<{{c}_{n_{k}3k}}$\\
15: \ \ \ \  \ \ \ \ \ \ \ $n^{*}=n_{k}$; $s^{*}=3$.\\
16: \ \ \ \  \ \ \ \ \textbf{EndIf} \\
17: \ \ \ \  \ \ \ $x_{n^{*}s^{*}k}=1$.\\
18: \ \ \ \textbf{EndIf}\\
19: \ \ \ \textbf{If} $k\in {{\mathcal{K}}_{t}}$\\
20: \ \ \ \  \ \ \ Select the subband 1 of some BS or the subband 2 of some\\
21: \ \ \ \  \ \ \ PBS to achieve the maximal utility: ${{c}_{n^{*}s^{*}k}}-{{\mu }_{n^{*}s^{*}}}$.\\
22: \ \ \ \  \ \ \ $x_{n^{*}s^{*}k}=1$.\\
23: \ \ \ \textbf{EndIf}\\
24: \ \ \ Feedback association information ${{x}_{{{n}^{*}}{{s}^{*}}k}}=1$ to the BS ${{n}^{*}}$.\\
25: \ \textbf{EndIf } \\ \bottomrule[0.5pt]
\end{tabular}
\label{alg_1}
\end{table}
\par
In the subproblem \eqref{eq17}, the optimal load ${{y}_{n1}}$ of BS $n\in {{\mathcal{N}}_{mp}}$ on subband 1 can be calculated according to Karush-Kuhn-Tucker (KKT) \cite{27} conditions and given by
\begin{equation}\label{eq22}
y_{n1}^{t+1}=\exp \left( \mu _{n1}^{t}-1 \right).
\end{equation}
\par
Similarly, the optimal load ${{y}_{n2}}$ of BS $n\in {{\mathcal{N}}_{p}}$ on subband 2 can be given by
\begin{equation}\label{eq23}
y_{n2}^{t+1}=\exp \left( \mu _{n2}^{t}-1 \right).
\end{equation}
\par
After getting the optimal load at time slot $t$, we can update the multiplier ${{\mu }_{n1}}$ of BS $n\in {{\mathcal{N}}_{mp}}$ on subband 1 using
\begin{equation}\label{eq24}
\mu _{n1}^{t+1}=\mu _{n1}^{t}-{\xi }^{t} \left( y_{n1}^{t}-\sum\limits_{k\in \mathcal{K}}{x_{n1k}^{t}} \right),
\end{equation}
Similarly, the multiplier ${{\mu }_{n2}}$ of BS $n\in {{\mathcal{N}}_{p}}$ on subband 2 can be updated by
\begin{equation}\label{eq25}
\mu _{n2}^{t+1}=\mu _{n2}^{t}-{\xi }^{t} \left( y_{n2}^{t}-\sum\limits_{k\in \mathcal{K}}{x_{n2k}^{t}} \right),
\end{equation}
where ${\xi }^{t}$ is a small enough stepsize for updating ${{\mu }_{ns}}$ at at time slot $t$. To this end, we can adopt Bertsekas's stepsize rule, i.e., equation (6.60) in \cite{28}. Moreover, Shen et al. \cite{17} propose a dual coordinate method to find optimal $\boldsymbol{\mu }$. These approaches ensure a faster convergence rate for the whole algorithm, but it may occupy a higher calculation complexity. For simplicity, we just consider a constant stepsize (special case for Bertsekas's stepsize rule). As for other rules, we can easily apply them into the proposed algorithm, and thus we will no longer take them into account.
\par
Evidently, the updating procedure of ${{y}_{ns}}$ and ${{\mu }_{ns}}$ takes place on BS's side, which is listed in Algorithm 2. When the BS $n$ is MBS, it updates the load $y_{n1}^{t}$ and $\mu _{n1}^{t+1}$ using equations \eqref{eq22} and \eqref{eq24} respectively, and this procedure performs in steps 1-9 of Algorithm 2; When the BS $n$ is PBS and adopts subband 1, it updates the load $y_{n1}^{t}$ and $\mu _{n1}^{t+1}$ using equations \eqref{eq22} and \eqref{eq24} respectively, and this procedure performs in steps 16-18 of Algorithm 2; When the BS $n$ is PBS and adopts subband 2, it updates the load $y_{n2}^{t}$ and $\mu _{n2}^{t+1}$ using equations \eqref{eq23} and \eqref{eq25} respectively, and this procedure performs in steps 20-22 of Algorithm 2.
\begin{table}[h]
\centering
\begin{tabular}{ll}
\toprule[1pt]
\textbf{Algorithm 2 at Base Station $n$} \\ \midrule[0.5pt]
1: \textbf{If } $n\in {{\mathcal{N}}_{m}}$ \\
2: \ \ \ \textbf{If } $t=0$ \\
3: \ \ \ \ \ \ Initialize stepsize ${{\xi }^{t}}$ and $\mu _{n1}^{t}$.\\
4: \ \ \ \textbf{Else } \\
5: \ \ \ \ \ \ Receive the information $x_{n1k}^{t}=1$ from any user $k\in \mathcal{K}$.\\
6: \ \ \ \ \ \ Calculate $y_{n1}^{t}$ using \eqref{eq22} and update $\mu _{n1}^{t+1}$ using \eqref{eq24}.\\
7: \ \ \ \ \ \ Broadcast information $\mu _{n1}^{t+1}$ to all users.\\
8: \ \ \ \textbf{EndIf } \\
9: \textbf{EndIf } \\
10: \textbf{If } $n\in {{\mathcal{N}}_{p}}$ \\
11: \ \ \ \textbf{For } $s\in {{\mathcal{S}}}$ \\
12: \ \ \ \ \ \ \ \textbf{If } $t=0$ \\
13: \ \ \ \ \ \ \ \ \ \ Initialize stepsize ${{\xi }^{t}}$ and $\mu _{ns}^{t}$.\\
14: \ \ \ \ \ \ \textbf{Else } \\
15: \ \ \ \ \ \ \ \ \ \ \textbf{If } $s=1$ \\
16: \ \ \ \ \ \ \ \ \ \ \ \ \ Receive the information $x_{n1k}^{t}=1$ from any user $k\in \mathcal{K}$.\\
17: \ \ \ \ \ \ \ \ \ \ \ \ \ Calculate $y_{n1}^{t}$ using \eqref{eq22} and update $\mu _{n1}^{t+1}$ using \eqref{eq24}.\\
18: \ \ \ \ \ \ \ \ \ \ \ \ \ Broadcast information $\mu _{n1}^{t+1}$ to all users.\\
19: \ \ \ \ \ \ \ \ \ \textbf{Else } \\
20: \ \ \ \ \ \ \ \ \ \ \ \ \ Receive the information $x_{n2k}^{t}=1$ from any user $k\in \mathcal{K}$.\\
21: \ \ \ \ \ \ \ \ \ \ \ \ \ Calculate $y_{n2}^{t}$ using \eqref{eq23} and update $\mu _{n2}^{t+1}$ using \eqref{eq25}.\\
22: \ \ \ \ \ \ \ \ \ \ \ \ \ Broadcast information $\mu _{n2}^{t+1}$ to all users.\\
23: \ \ \ \ \ \ \ \ \ \ \textbf{EndIf } \\
24: \ \ \ \ \ \ \ \textbf{EndIf } \\
25: \ \ \ \textbf{EndFor }\\
26: \textbf{EndIf }\\ \bottomrule[0.5pt]
\end{tabular}
\label{alg2}
\end{table}
\par
In the equations \eqref{eq24} and \eqref{eq25}, there are some interesting meanings. Specially, the multiplier $\mu_{n1}$ can be regarded as a message between any user $k\in \mathcal{K}$ and BS $n\in {{\mathcal{N}}_{mp}}$ on subband 1. Furthermore, it can be also seen as the service cost of BS $n\in {{\mathcal{N}}_{mp}}$ on subband 1, which should be dependent on the load distribution. When $\sum\nolimits_{k\in \mathcal{K}}{{{x}_{n1k}}}$ and ${{y}_{n1}}$ are deemed to be the serving demand and available service of BS $n\in {{\mathcal{N}}_{mp}}$ on subband 1 respectively, the cost $\mu_{n1}$ can tradeoff supply and demand. Consequently, the cost $\mu_{n1}$ will go up if the demand $\sum\nolimits_{k\in \mathcal{K}}{{{x}_{n1k}}}$ exceeds the supply ${{y}_{n1}}$ and vice versa. Similarly, the multiplier $\mu_{n2}$ represents a message between any user $k\in \mathcal{K}$ and BS $n\in {{\mathcal{N}}_{p}}$ on subband 2, and meanwhile it can be also interpreted as the service cost of BS $n\in {{\mathcal{N}}_{p}}$ on subband 2. When $\sum\nolimits_{k\in \mathcal{K}}{{{x}_{n2k}}}$ and ${{y}_{n2}}$ are denoted as the serving demand and available service of BS $n\in {{\mathcal{N}}_{p}}$ on subband 2 respectively, the cost $\mu_{n2}$ can tradeoff supply and demand. Thus, the cost $\mu_{n2}$ will go up if the demand $\sum\nolimits_{k\in \mathcal{K}}{{{x}_{n2k}}}$ exceeds the supply ${{y}_{n2}}$ and vice versa. In the whole association procedure, some users may not be associated with some overloaded BS when the latter increases its service price, but may be connected to some underloaded BS with decreasing price.
\par
The aforementioned two algorithms are listed in the pattern \cite{29}, which can give us a clear insight on exchanged information and overhead. The whole association procedure should interactively execute user's algorithm (Algorithm 1) and BS's algorithm (Algorithm 2). To this end, we give the following theorem.
\par
\textbf{Theorem 1.} If there exists $\varphi>0$ and ${{I}^{*}}>-\infty $, then
\par
\begin{equation}\label{eq26}
\underset{t}{\mathop{\inf }}\,I\left( {{\boldsymbol{\mu} }^{t}} \right)\le {{G}^{*}}+\varphi,
\end{equation}
where ${{I}^{*}}$ denotes the optimal value of problem \eqref{eq15}.
\par
\emph{Proof}:  The derivative of function $I\left(\boldsymbol{\mu}  \right)$ is calculated by
\begin{equation}\label{eq27}
\frac{\partial I}{\partial {{\mu }_{n1}}}\left(\boldsymbol{\mu} \right)={{y}_{n1}}\left( \boldsymbol{\mu}  \right)-\sum\limits_{k\in \mathcal{K}}{{{x}_{n1k}}\left( \boldsymbol{\mu}  \right)},
\end{equation}
and
\begin{equation}\label{eq28}
\frac{\partial I}{\partial {{\mu }_{n2}}}\left(\boldsymbol{\mu} \right)={{y}_{n2}}\left( \boldsymbol{\mu}  \right)-\sum\limits_{k\in \mathcal{K}}{{{x}_{n2k}}\left( \boldsymbol{\mu}  \right)}.
\end{equation}
\par
Since the number of users scattered in network is limited in real life, $\sum\nolimits_{k\in \mathcal{K}}{{x}_{n1k}}$, $\sum\nolimits_{k\in \mathcal{K}}{{x}_{n2k}}$, ${{y}_{n1}}$ and ${{y}_{n2}}$ are bounded. Thus, the function $\partial I$ is bounded. Evidently, our problem meets the necessary conditions of proposition 6.3.6 in \cite{13}, and the theorem can be proved by applying this proposition.
\par
Next, we will give some analyses for the algorithm complexity. As shown in Algorithm 1, any user has a computation complexity of $\mathcal{O}\left( {{N}_{mp}}+{{N}_{p}}\right)$. In the Algorithm 2, any MBS has a computation complexity of $\mathcal{O}\left( {{K}}\right)$, and any PBS has the one of $\mathcal{O}\left( SK \right)$. Thus, we can easily know that all users have the computation complexity of $\mathcal{O}\left(  {{N}_{mp}}{{K}}+{{N}_{p}}{{K}} \right)$, all BSs has the one of $\mathcal{O}\left(N_{mp}SK\right)$. Thus, when a centralized algorithm is adopted, the computation complexity may be $\mathcal{O}\left(  {{N}_{mp}}{{K}}+{{N}_{p}}{{K}} \right)$ at each iteration. Evidently, a centralized algorithm for solving the proposed problem should be more complicated than the advocated algorithm.
\par
Moreover, the equations \eqref{eq24} and \eqref{eq25} show that BSs only require very little local information to adjust the multiplier $\boldsymbol{\mu }$ in a completely distributed manner. Specially, each BS broadcasts its service cost that only contains very little information to all users, and each user should send its service demand to the BS to which it expects to connect. Evidently, the amount of exchanged information of the proposed algorithm should be ${N_{m}}+{N_{p}}S+K$ at each iteration. Unlike the proposed algorithm, a centralized algorithm may have the amount of exchanged information that is proportional to $N_{mp}SK$ at each iteration. Thus, the proposed algorithm should be more practical and favored for some cases, especially for large-scale problems.
\section{Numerical results}
In D2D-enabled HCNs, the transmit power of MBSs, PBSs and D2D TXs is 46 dBm, 30 dBm and 20 dBm respectively. In the D2D-enabled HCNs, MBSs are deployed into a conventional cellular network, potential D2D pairs, PBSs and cellular users are uniformly and independently scattered into each macrocell. We assume that the distance between MBSs is 1000 m, and the distance between D2D RX and D2D TX is greater than 10 m and less than 50 m. As for the propagation environment, we regard $PL\left( d \right)=128.1+37.6\log 10\left( d \right)$ dB as the pathloss model of MBS, and see $PL\left( d \right)=140.7+36.7\log 10\left( d \right)$ dB as the one of PBS and D2D pairs \cite{30}. In these pathloss models, parameter $d$ represents the distance between the receiver and transmitter in kilometers. Moreover, MBSs and PBSs own log-normal shadowing with standard deviation 10 dB, and D2D pairs have log-normal shadowing with standard deviation 12 dB. In the D2D-enabled HCNs, the noise power spectral density equals to -174dBm/Hz.
\par
Considering that the proposed association scheme maximizes the network-wide utility, we can simply call it Max-Utility association. To highlight the effectiveness of this association scheme, we introduce other association schemes for comparison, which mainly include two types of association schemes, i.e., conventional association and load balancing association. Specially, the former refers to Max-Rate association and Max-SINR association, and the latter includes Rate Bias association and SINR Bias association. Evidently, the Max-Rate association and Rate Bias association are closely related to available bandwidth, but the Max-SINR association and SINR Bias are not. The detailed descriptions for them are listed as follows.
\par
\textbf{Maximal Rate (Max-Rate) Association:} In the Max-Rate association, we replace the utility ${{c}_{n^{*}s^{*}k}}-{{\mu }_{n^{*}s^{*}}}$ in Algorithm 1 by the achievable rate ${{r}_{n^{*}s^{*}k}}$ and meanwhile replace ${{c}_{n_{k}3k}}$ by ${{r}_{n_{k}3k}}$, then perform the steps 6-23 of Algorithm 1.
\par
\textbf{Maximal SINR (Max-SINR) Association:} In the Max-SINR association, we replace the utility ${{c}_{n^{*}s^{*}k}}-{{\mu }_{n^{*}s^{*}}}$ in Algorithm 1 by the achievable rate ${{\text{SINR}}_{n^{*}s^{*}k}}$ and meanwhile replace ${{c}_{n_{k}3k}}$ by ${{\text{SINR}}_{n_{k}3k}}$, then perform the steps 6-23 of Algorithm 1.
\par
\textbf{Rate Bias Association:} In the Rate Bias association, we replace the utility ${{c}_{n^{*}s^{*}k}}-{{\mu }_{n^{*}s^{*}}}$ in Algorithm 1 by the achievable rate ${{r}_{n^{*}s^{*}k}}{e}^{-\mu _{n^{*}s^{*}}}$ and meanwhile replace ${{c}_{n_{k}3k}}$ by ${{r}_{n_{k}3k}}$, then perform the steps 6-23 of Algorithm 1. Note that $\mu _{ns}^{*}$ is the optimal solution obtained by the proposed algorithm. Evidently, we can easily find that this association scheme should be equivalent to the proposed one (Max-Utility association) according to the association rule of Algorithm 1.
\par
\textbf{SINR Bias Association:} In the SINR Bias association, we replace the utility ${{c}_{n^{*}s^{*}k}}-{{\mu }_{n^{*}s^{*}}}$ in Algorithm 1 by the achievable rate ${\text{SINR}_{n^{*}s^{*}k}}/{{p}_{n^{*}s^{*}}}$ and meanwhile replace ${{c}_{n_{k}3k}}$ by ${\text{SINR}_{n_{k}3k}}/{{p}_{n_{k}3}}$, then perform the steps 6-23 of Algorithm 1. Note that ${{p}_{ns}}$ is the transmit power of transmitter $n$ on subband $s$, and the subband $s$ will be not considered in the association procedure when ${{p}_{ns}}=0$ mW.
\par
As for association performance, we mainly focus on the load balancing level, the cumulative distribution function (CDF) of effective rates, the coverage probability of effective rates and convergence of proposed algorithm.
\subsection{Load Balancing Level}
\begin{figure}[!t]
\centering
\centerline{\includegraphics[width=4in]{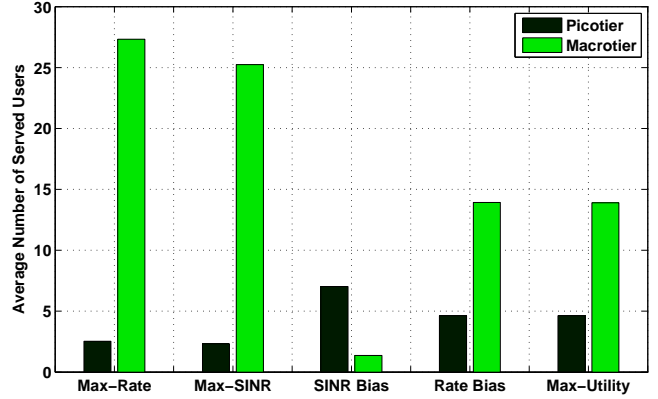}}
\caption{The average numbers of users served by per tier for different association schemes.}
\label{fig3}
\end{figure}
\par
Fig. \ref{fig3} shows the load distributions for different association schemes. The associations Max-SINR and Max-Rate result in very unbalanced load distributions: most users are associated with the macrotier consisting of MBSs, and very few users can be served by the picotier consisting of PBSs. That's because the MBS has higher transmit power than PBS, and thus users associated with MBSs often have higher SINRs/rates than them associated with PBSs. The SINR Bias association achieves a relatively high balancing level: more users favour picotier since they may achieve higher SINRs on subband 2 of PBSs than the ones on subband 1 of MBSs. As illustrated in Fig. \ref{fig3}, the associations Max-Utility and Rate Bias can balance the loads among different network tiers and achieve almost the same effect. These schemes can offload the low-rate (achievable rate) users associated with overloaded MBSs in the associations Max-SINR and Max-Rate to the adjacent underloaded BS.
\par
To measure the status of the system load balancing level in a more refined metric, we introduce the Jain's fairness index \cite{31}, which is given by
 \begin{equation}\label{eq29}
J=\frac{{{\left( \sum\nolimits_{n\in {{\mathcal{N}}_{mp}}}{{{y}_{n}}} \right)}^{2}}}{{{N}_{mp}}\sum\nolimits_{n\in {{\mathcal{N}}_{mp}}}{y_{n}^{2}}},
\end{equation}
where $\sum\nolimits_{k\in \mathcal{K}}\sum\nolimits_{s\in \mathcal{S}}{{{x}_{nsk}}}={{y}_{n}}$ represents the load of BS ${n}$. The larger $J$ that belongs to the interval $\left[ {1}/{N_{mp}},1 \right]$ means more balanced load distribution among the given cells. Thus, the Jain's fairness index in this paper can be also named as the load balancing index (LBI). Significantly, we just consider the load balancing level of all BSs, which doesn't refer to D2D TXs. Since any D2D TX just has one or zero served user (its corresponding D2D RX), the load balancing level of D2D TX doesn't need to be considered.
\begin{figure}[!t]
\centering
\centerline{\includegraphics[width=4in]{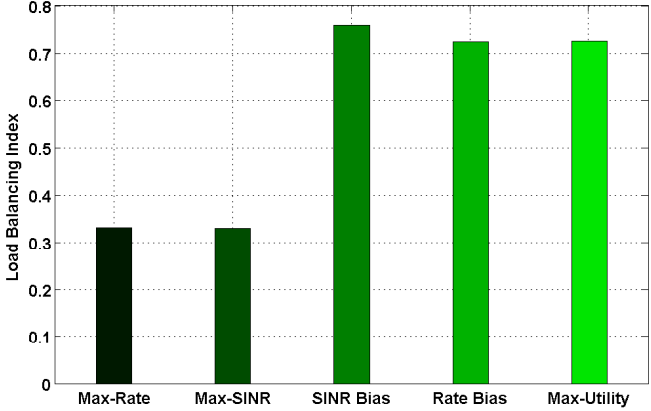}}
\caption{The load balancing indices for different association schemes.}
\label{fig4}
\end{figure}
\par
Fig. \ref{fig4} shows the load balancing indices for different association schemes. Since the SINR Bias association doesn't depend on transmit power of BSs and performs user association in a very random manner, it achieves the highest LBI among all schemes. Unlike the SINR Bias association, the associations Max-Rate and Max-SINR should achieve the lowest LBI. As shown in this figure, the associations Max-Utility and Rate Bias achieve a higher LBI than associations Max-Rate and Max-SINR because of their offloading capabilities.
\begin{figure}[!t]
\centering
\centerline{\includegraphics[width=4in]{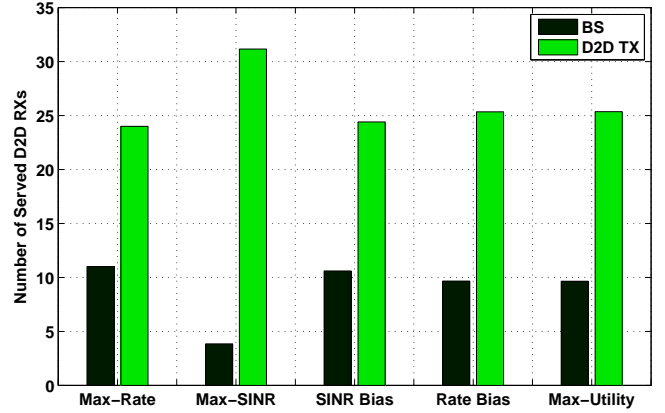}}
\caption{The numbers of D2D RXs served by BSs or D2D TXs for different association schemes.}
\label{fig5}
\end{figure}
\par
Fig. \ref{fig5} shows the numbers of D2D RXs served by BSs or D2D TXs for different association schemes. The Max-SINR association supports the most D2D pairs among all schemes, and the associations Max-Utility and Rate Bias support more D2D pairs than the associations Max-Rate and SINR Bias. When the number of D2D pairs is relatively small, the D2D RXs can receive the weaker interference from D2D TXs than from BSs. Thus, more D2D RXs can be served by the corresponding D2D TXs. However, due to the limited bandwidth, the achievable rates of D2D TXs may have not enough superiority, which leads to more fewer D2D pairs are supported. It is noteworthy that the numbers of D2D pairs supported by Max-Rate and by Max-SINR will be decrease with the number of D2D pairs due to the stronger and stronger interference. In the SINR Bias association, more D2D RXs select the corresponding D2D TXs because of large shadowing fading. Unlike other schemes, the associations Max-Utility and Rate Bias trade off load and achievable rate, i.e., reduce the achievable rate by offloading. Thus, this operation will be beneficial to the utilization of D2D pairs. Significantly, the capabilities of supporting D2D pairs represents their offloading capabilities. As we know, more balanced load distribution is often beneficial to the full utilization of network resources and improving user experience.
\subsection{Load Balancing Gain}
The load balancing gain represents that the association scheme improves user experience by balancing loads among different BSs.
\begin{figure}[!t]
\centering
\centerline{\includegraphics[width=4in]{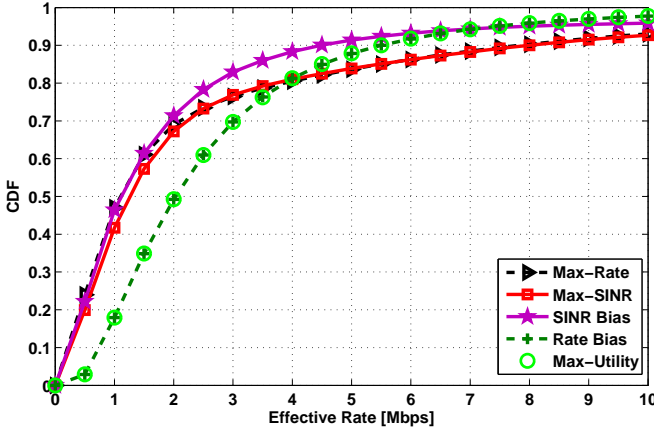}}
\caption{The CDF of effective rates for four different association schemes.}
\label{fig6}
\end{figure}
\par
Fig. \ref{fig6} plots the cumulative distribution function (CDF) of effective rates for different association schemes. As illustrated in Fig. \ref{fig6}, the associations Max-Utility and Rate Bias own the fewest low-rate (effective rate) users among all schemes, while the SINR Bias association has the most low-rate users among all schemes. Although the SINR Bias association has an offloading ability, it cannot guarantee that the offloading operation can improve user experience because of highly random association. Unlike the SINR Bias association, the associations Max-Utility and Rate Bias can offload the low-rate (achievable rate) users associated with overloaded MBSs in the associations Max-SINR and Max-Rate to the adjacent underloaded BS, which can improve user experience.
\begin{figure}[!t]
\centering
\centerline{\includegraphics[width=4in]{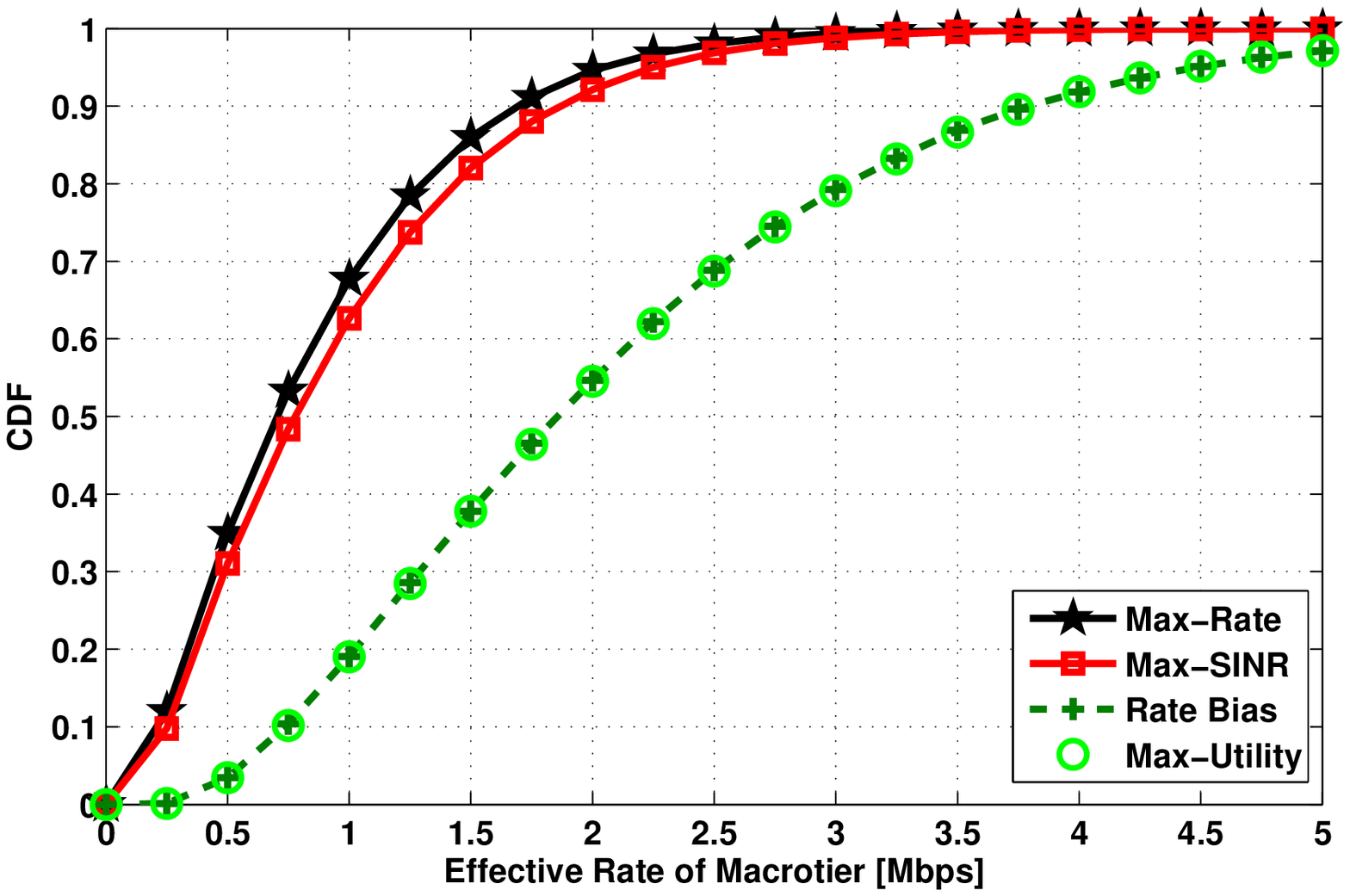}}
\caption{The CDF of effective rates of macrotier for different association schemes.}
\label{fig7}
\end{figure}
\par
Fig. \ref{fig7} plots the CDF of effective rates of macrotier for different association schemes. As mentioned in previous section, the associations Max-Utility and Rate Bias can offload the low-rate (achievable rate) users associated with overloaded MBSs in the associations Max-SINR and Max-Rate to the adjacent underloaded BS. Evidently, this operation can improve user experience and thus the associations Max-Utility and Rate Bias have fewer low-rate users than the associations Max-SINR and Max-Rate. Since the Max-SINR association doesn't consider available bandwidth but the Max-Rate association does it, the latter has more low-rate users.
\par
Evidently, the associations Max-Utility and Rate Bias achieve more higher load balancing gain than other schemes.
\subsection{Resource Partitioning Gain}
The resource partitioning gain represents that the association scheme improves user experience by partitioning resource.
\begin{figure}[!t]
\centering
\centerline{\includegraphics[width=4in]{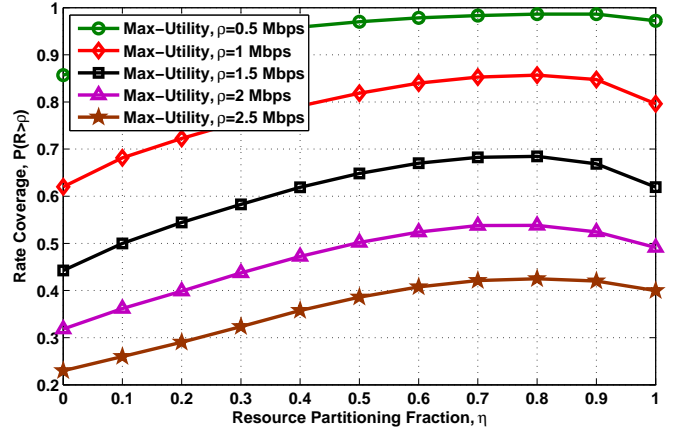}}
\caption{The coverage probability of effective rates for four different target rates.}
\label{fig8}
\end{figure}
\par
Fig. \ref{fig8} shows the coverage probability of effective rates for different target rates. Note that the rate coverage represents the proportion of the users whose effective rates are greater than target rate $\rho$ in all users. Seen from Fig. \ref{fig8}, we can find that the coverage probability initially increases with increasing $\eta$ and then decreases with it. That's because the offloaded users may receive weaker and weaker interference from MBSs as this fraction increases, but then their effective rates should decrease with increasing fraction due to fewer available bandwidth. Through a direct observation, we can easily find the resource partitioning gain, i.e., the association scheme with resource partitioning provides a higher coverage probability than the one without resource partitioning. Moreover, the coverage probability should be higher and higher as the target rate becomes lower and lower.
\subsection{Convergence}
\begin{figure}[!t]
\centering
\centerline{\includegraphics[width=4in]{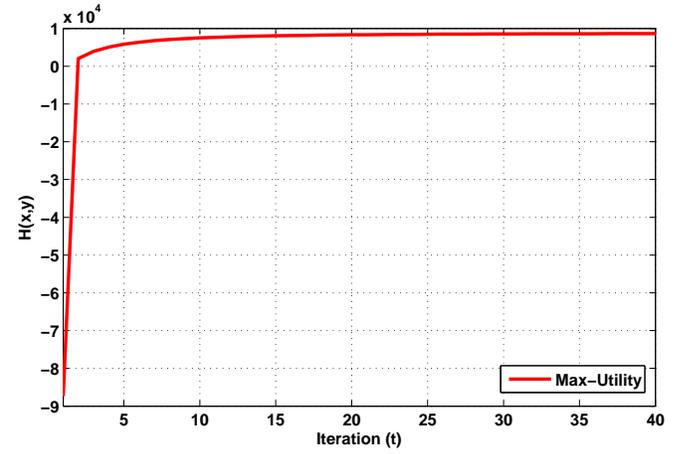}}
\caption{The convergent sum-utility of the Max-Utility association algorithm.}
\label{fig9}
\end{figure}
\par
Fig. \ref{fig9} plots the convergent sum-utility of the Max-Utility association algorithm, where parameter $t$ is $t\text{-th}$ iteration. To achieve optimal solutions and implementation simplicity, we only consider a constant stepsize for updating multiplier $\boldsymbol{\mu }$ in the proposed algorithm. As illustrated in Fig. \ref{fig9}, the proposed algorithm can converge in just a few iterations, which means it can be well applied in reality, especially in large-scale case.
\section{Conclusion}
For the D2D-enabled HCNs, we propose an offloading scheme with maximizing network-wide utility, and then design a highly effective distributed algorithm by dual decomposition. Numerical results show that the proposed association scheme can provide a load balancing gain, and meanwhile reveals the offloading capacity of D2D pairs. Moreover, the proposed resource partitioning scheme can also provide its gain. Future work can include designing a dynamic association algorithm, introducing power control and considering uplink scenario.
\section*{Acknowledgment}
This work was supported by the National Natural Science Foundation of China under Grants 61372101, 61422105 and 61271018, the National Science and Technology Major Project of China under Grants 2013ZX03003006-002 and 2012ZX03004-005-003, the Natural Science Foundation of Jiangsu Province under Grant BK20130019, the Research Project of Jiangsu Province under Grant BE2012167, and the Program for New Century Excellent Talents in University under Grant NCET-11-0088.


\end{document}